\documentclass[11pt,a4paper]{article}

\usepackage[utf8]{inputenc}
\usepackage[T1]{fontenc}
\usepackage[english]{babel}

\usepackage{fullpage}
\usepackage{setspace}
\usepackage{geometry}
\usepackage{amsmath, amssymb, amsfonts, amsthm}
\usepackage{amsopn}
\theoremstyle{plain}

\usepackage{graphicx}   % Add [draft] for fast Overleaf compile; remove for final PDF
\usepackage{subcaption}

\usepackage{tikz}
\usetikzlibrary{arrows, backgrounds, decorations, trees, shapes.geometric}

\usepackage{booktabs, tabularx}
\usepackage{multirow}
\usepackage{array, longtable}
\usepackage{tabularray}
\usepackage{rotating}
\usepackage[flushleft]{threeparttable}
\usepackage{arydshln}
\newcolumntype{L}[1]{>{\raggedright\let\newline\\\arraybackslash\hspace{0pt}}m{#1}}
\newcolumntype{C}[1]{>{\centering\let\newline\\\arraybackslash\hspace{0pt}}m{#1}}
\newcolumntype{R}[1]{>{\raggedleft\let\newline\\\arraybackslash\hspace{0pt}}m{#1}}

\usepackage[table]{xcolor}
\definecolor{Peakred}{RGB}{255,69,0}
\definecolor{Peakblue}{RGB}{55,126,184}
\definecolor{Peakgreen}{RGB}{77,175,74}

\usepackage{pdflscape}
\usepackage{lscape}
\usepackage{enumitem}
\usepackage{comment}
\usepackage[toc,page]{appendix}
\usepackage[capposition=top]{floatrow}
\usepackage[bottom]{footmisc}

\usepackage{natbib}

\usepackage[colorlinks=true, linkcolor=blue, urlcolor=blue,
            citecolor=blue, anchorcolor=blue]{hyperref}

\title{Biodiversity Media Narratives and Stock Market Performance: Evidence from Europe}
\author{
Andr\'{e}s Azqueta-Gav\'{a}ld\'{o}n\thanks{Independent Researcher. 
Email: andres.azqueta@gmail.com} 
\and 
Ben Jabeur Sami\thanks{ESDES, UCLy, Lyon. 
Email: sbenjabeur@univ-catholyon.fr} 
\and 
Leila Hedhili\thanks{University of Tunis. 
Email: leila.hedhili@isg.u-tunis.tn}
}
\date{\today}

\begin{document}
\maketitle

% =========================
% ABSTRACT
% =========================
\begin{abstract}
This study constructs novel biodiversity related media risk indicators for France, Germany, Italy, and Spain over 2015--2025, capturing media attention to biodiversity threats using the GDELT Global Knowledge Graph. Using panel Granger causality tests and an augmented inverse probability weighting (AIPW) event-study design, we find highly significant evidence that biodiversity risk reduces stock prices, with effects peaking between 3 and 10 months after a shock. Moreover, we uncover a marked asymmetry whereby the positive effects of low biodiversity risk episodes outweigh the negative effects of high-risk episodes. Results are robust across quantiles of the return distribution and hold when controlling for European equity market volatility and economic policy uncertainty. Our findings provide the first evidence that biodiversity media narratives drive stock market valuations in Europe.

% G12 - Asset Pricing
% G14 - Information and Market Efficiency
% Q57 - Ecological Economics
% C43 - Index Numbers and Aggregation
% C55 - Large Data Sets: Modeling and Analysis
\end{abstract}

\clearpage

% =========================
% INTRODUCTION
% =========================
\section{Introduction}
\label{Introduction}

Biodiversity loss has emerged as a critical global challenge alongside climate change,
with the Intergovernmental Science-Policy Platform on Biodiversity and Ecosystem
Services (IPBES) documenting unprecedented species extinction
rates~\citep{ipbes2019}. Despite growing recognition of biodiversity's economic
importance---ecosystem services contribute an estimated \$125--140 trillion annually
to global GDP \citep{costanza2014}---its impact on financial markets remains
underexplored. While climate risk has become central to asset pricing
research~\citep{bolton2021}, biodiversity risk lacks systematic measurement and
financial market integration~\citep{giglio2025}.

Several studies find that the media is a powerful tool for increasing public attention
about environmental concerns, stock market activity, trading behavior, and investor
sentiment~\citep{ardia2023climate, filippou2024media, peress2014media}. Information
provided by the news media about biodiversity risks can drive economic outcomes, as
documented by~\citet{shiller2017narrative}. Research on narrative biodiversity risk in
finance is relatively new. \citet{giglio2025} construct a monthly biodiversity risk
index based on coverage in the \textit{New York Times} from 2010 to 2023 and find
that biodiversity risk impacts equity prices. In the same vein, \citet{ma2024biodiversity}
propose a biodiversity concern index for China and reveal a significant effect on stock
returns. However, for European countries, no biodiversity-related media narratives have
yet been proposed, despite the region facing major challenges linked to biodiversity
loss and ecosystem collapse~\citep{garel2025nature}.

This paper makes three contributions. First, we develop text-based biodiversity risk
indicators for France, Germany, Italy, and Spain using the GDELT Global Knowledge
Graph, capturing media attention to biodiversity threats from 2015 to the present. Our
methodology follows established word2vec biodiversity keyword
approaches~\citep{sautner2023} adapted to environmental contexts. Second, we document
that biodiversity risk negatively predicts stock market returns, providing the first
evidence that biodiversity concerns affect aggregate market valuations in Europe.
Third, we show that these effects are heterogeneous across quantiles of the return
distribution, with the strongest impact concentrated in the tails---precisely where
systemic risk is most relevant for investors and regulators.

Our findings contribute to the growing literature on environmental risk
pricing~\citep{hong2019, engle2020} and have implications for sustainable finance
frameworks including the EU Taxonomy and the Task Force on Nature-related Financial
Disclosures (TNFD).

% =========================
% DATA AND METHODOLOGY
% =========================
\section{Data and Methodology}
\label{sec_data_methodology}

\subsection{Biodiversity Risk Indicators}

We construct monthly biodiversity risk indicators using the GDELT Global Knowledge
Graph (GKG) \citep{leetaru2013}, a comprehensive open-source database that
systematically monitors international news media, extracting structured metadata
including source identifiers, document URLs, and algorithmically-assigned thematic
tags. Our analysis focuses on GDELT~2.0 (2015--present), which offers enhanced
coverage of international sources. Following established word2vec methodologies for
identifying semantically-related terms~\citep{sautner2023, mikolov2013}, we employ
100+ biodiversity keywords across four categories: core biodiversity concepts,
ecosystem types and habitats, species and conservation, and threats and degradation.
For each country, we implement a two-stage geographic and content filtering procedure
to retain only articles discussing biodiversity issues specifically related to the focal
country. Full details on the sampling strategy, keyword taxonomy, and geographic
filtering are provided in Appendix~\ref{app:data}.

Figure~\ref{fig:biodiversity_indicators} presents three complementary views of the
resulting indicators across France, Germany, Italy, and Spain over 2015--2025. The top
panel shows raw monthly article counts. France consistently records the highest media
coverage, with a peak of approximately 225 articles in early 2017, followed by Germany
which reached around 210 articles in 2016. Italy and Spain display lower and broadly
similar coverage levels. Across all countries, raw mentions decline after 2016--2017
and stabilize at lower levels. The middle panel displays the standardized indicators,
which correct for differences in baseline coverage across countries. The shaded area
around 2015--2016 denotes the initialization period of GDELT~2.0, during which
coverage was ramping up and indicators should be interpreted with caution. All four
indicators exhibit episodic spikes, with Italy recording the highest standardized peak
(approximately 6 standard deviations) in 2016. From 2019 onwards, the indicators
converge toward zero with reduced volatility. The bottom panel reports the
cross-country correlation matrix. All pairwise correlations are positive and moderate,
ranging from 0.39 (Germany--Italy) to 0.69 (France--Germany), consistent with common
European-level biodiversity narratives while preserving meaningful country-specific
variation. This validates the use of country-level indicators rather than a single
aggregate index.

\begin{figure}[!htbp]
\centering
\includegraphics[width=\linewidth,height=0.72\textheight,keepaspectratio]%
    {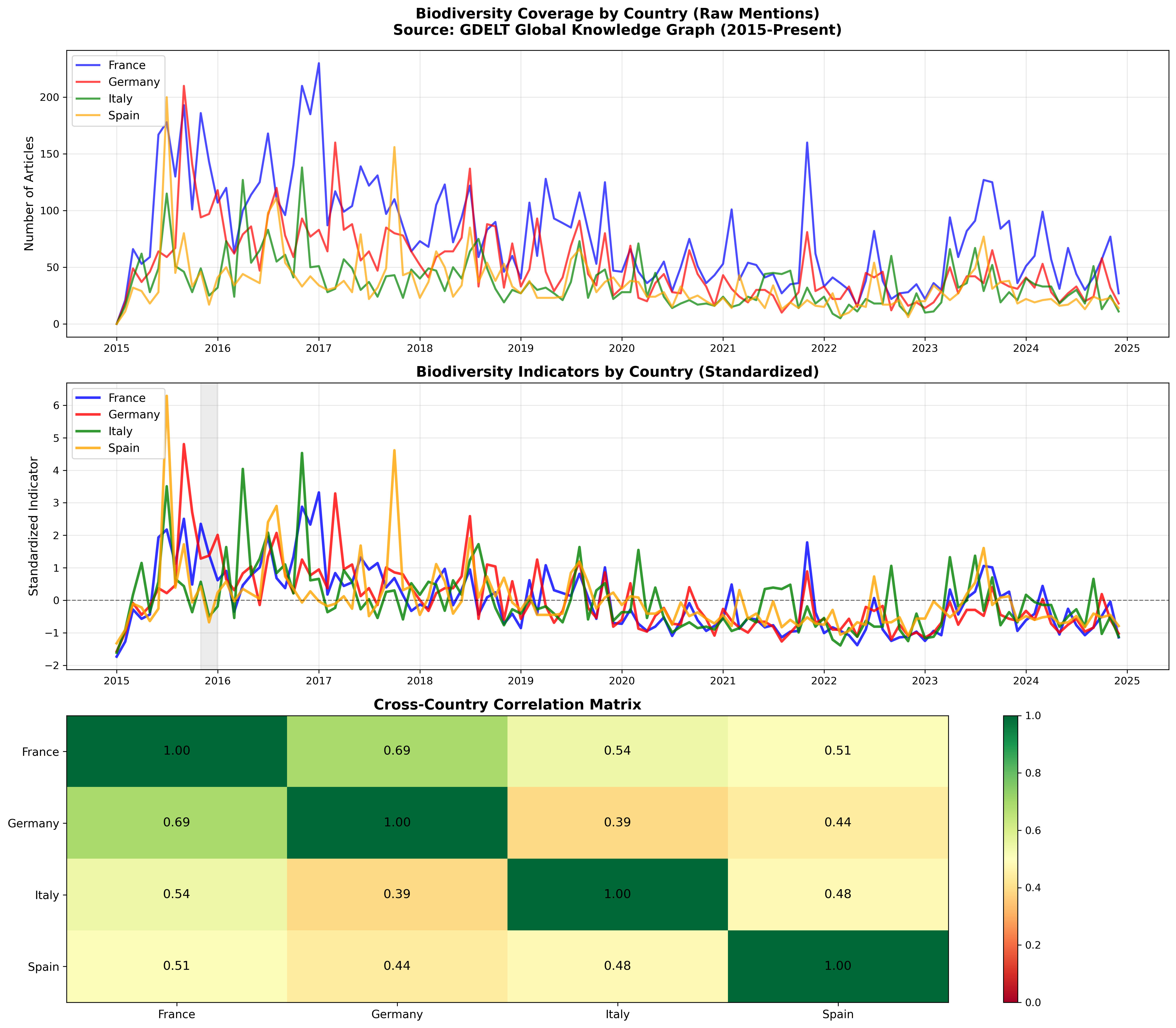}
\caption{Biodiversity Risk Indicators across European Countries, 2015--2025.
\textit{Note:} Top panel: raw monthly article counts from GDELT. Middle panel:
standardized indicators (shaded area denotes GDELT~2.0 initialization period).
Bottom panel: cross-country correlation matrix of standardized indicators.}
\label{fig:biodiversity_indicators}
\end{figure}

Table~\ref{tab:descriptive_stats} reports summary statistics for the main 
variables. The biodiversity risk index is standardized by construction 
(mean zero, unit variance, max $= 6.29$), while the stock price index 
averages 12,975 points (SD $= 7{,}136$), providing context for the 
treatment effect magnitudes reported in 
Section~\ref{sec_impact_stock_markets}.

\begin{table}[!htbp]
\centering
\caption{Descriptive Statistics for the EU4 Panel}
\label{tab:descriptive_stats}
\begin{tabular}{lccccc}
\toprule
\textbf{Variable} & \textbf{Mean} & \textbf{SD} & \textbf{Min} & \textbf{Max} & \textbf{N} \\
\midrule
Stock price index & 12{,}975.39 & 7{,}135.98 & 4{,}237.48 & 34{,}750.00 & 480 \\
Biodiversity risk index (standardized) & 0.00 & 1.00 & $-$1.73 & 6.29 & 480 \\
Euro Stoxx 50 index & 20.66 & 6.59 & 11.99 & 48.59 & 480 \\
Economic Policy Uncertainty (EPU) & 225.32 & 161.08 & 31.70 & 1095.93 & 480 \\
\bottomrule
\end{tabular}
\begin{flushleft}
\footnotesize
\textit{Notes:} The sample comprises monthly observations for France, Germany, Italy
and Spain over 2015--2025 ($N = 480$ country-month observations). The biodiversity
risk index is standardized to have mean zero and unit variance.
\end{flushleft}
\end{table}

% =========================
% IMPACT ON STOCK MARKETS
% =========================
\section{Impact on Stock Markets}
\label{sec_impact_stock_markets}

Stock market prices are measured using monthly closing values of the CAC~40 (France),
DAX (Germany), FTSE~MIB (Italy), and IBEX~35 (Spain), sourced from Refinitiv
Datastream over the 2015--2025 period.

\subsection{Panel Granger Causality}

We apply the panel Granger causality test proposed by~\citet{dumitrescu2012granger}
to assess the causal relationship between biodiversity shocks (standardized values)
and European stock market prices. Table~\ref{tab:granger_biodiversity} indicates that
biodiversity risk Granger-causes stock market prices (Z-stat $= 3.24$, $p < 0.01$),
while the reverse direction is not significant, confirming the unidirectional nature
of the relationship.\footnote{We verify that reverse causality does not arise
mechanically from the construction of our biodiversity indicator by running the test
using the raw standardized index ($z\_bio$) rather than any orthogonalized variant.
The absence of reverse causality therefore constitutes an independent empirical result
rather than a consequence of prior identification.} Similar results have been reported
for US equity markets by~\citet{giglio2025}.

\begin{table}[!htbp]
\centering
\caption{Panel Granger Causality between Biodiversity Risk and Stock Market Prices}
\label{tab:granger_biodiversity}
\begin{tabular}{lcc}
\toprule
\textbf{Direction of Causality} & \textbf{Z-tilde} & \textbf{p-value} \\
\midrule
Biodiversity $\rightarrow$ Stock Prices & 3.2442 & 0.0012$^{***}$ \\
Stock Prices $\rightarrow$ Biodiversity & $-$0.6291 & 0.5293 \\
\bottomrule
\end{tabular}
\begin{flushleft}
\footnotesize
\textit{Notes:} $^{*}$, $^{**}$, and $^{***}$ denote statistical significance at the
10\%, 5\%, and 1\% levels, respectively. Panel Granger causality test following
\citet{dumitrescu2012granger}. Both directions estimated using the raw standardized
biodiversity index ($z\_bio$) and monthly stock price closing values.
\end{flushleft}
\end{table}

\subsection{AIPW Event Study Design}

To assess the dynamic and nonlinear effects of biodiversity shocks across different
risk regimes, we estimate average treatment effects using the augmented inverse
probability weighting (AIPW) estimator of~\citet{robins1994}, implemented via
Stata's \texttt{teffects aipw} command \citep{statacorp2025}, following the exposition in~\citet{wooldridge2010} and recent advances in doubly robust estimation for panel and event study settings~\citep{santanna2020, arkhangelsky2024}. The AIPW estimator is doubly robust: it yields consistent estimates if either the propensity score model or the outcome regression is correctly specified~\citep{wooldridge2010}.

\textbf{Treatment definition.} For each country, we compute the 25th and 75th
percentiles of the standardized biodiversity risk index ($z\_bio$) \textit{within
country} and define three mutually exclusive regimes: (i)~a \textit{high
biodiversity risk} regime (top quartile); (ii)~a \textit{low biodiversity risk}
regime (bottom quartile); and (iii)~a \textit{middle regime} (interquartile range),
which serves as the reference category. Quartile-based thresholds are preferred over
more extreme cutoffs (e.g., 10th and 90th percentiles) because the sample comprises
monthly observations for only four countries over 2015--2025, and broader quantile
groups provide a sufficient number of treated observations to ensure stable
estimation. The estimator is run separately for high-risk versus middle observations
and for low-risk versus middle observations.

\textbf{Outcome regression.} For each horizon $h = 0, \ldots, 20$, the dependent
variable is the stock price index observed $h$ months ahead,
$y_{i,t+h} = \text{price}_{i,t+h}$. The outcome regression estimated in the
augmentation step takes the form:
\begin{equation}
y_{i,t+h} = \alpha_h + \beta_h D_{i,t} + \gamma_h \text{price}_{i,t-1} +
\varepsilon_{i,t+h},
\label{eq:outcome}
\end{equation}
where $D_{i,t}$ is alternatively the indicator for high or low biodiversity risk.
Accordingly, the estimated average treatment effects are expressed in index points.

\textbf{Propensity score model.} The conditional probability of treatment assignment
is estimated using a logit model:
\begin{equation}
\Pr(D_{i,t} = 1 \mid X_{i,t}) = \Lambda\!\left(\delta_0 + \delta_1
\text{price}_{i,t-1}\right),
\label{eq:pscore}
\end{equation}
where $\Lambda(\cdot)$ denotes the logistic cumulative distribution function and
$X_{i,t}$ contains the one-month lag of the stock price index. The estimated
propensity score is used to construct inverse probability weights, which are combined
with the outcome regression to obtain the doubly robust AIPW estimator. Standard
errors are clustered at the country level to account for within-country serial
dependence.

\subsection{Results}

\begin{figure}[!htbp]
\centering
\includegraphics[width=\linewidth,height=0.72\textheight,keepaspectratio]{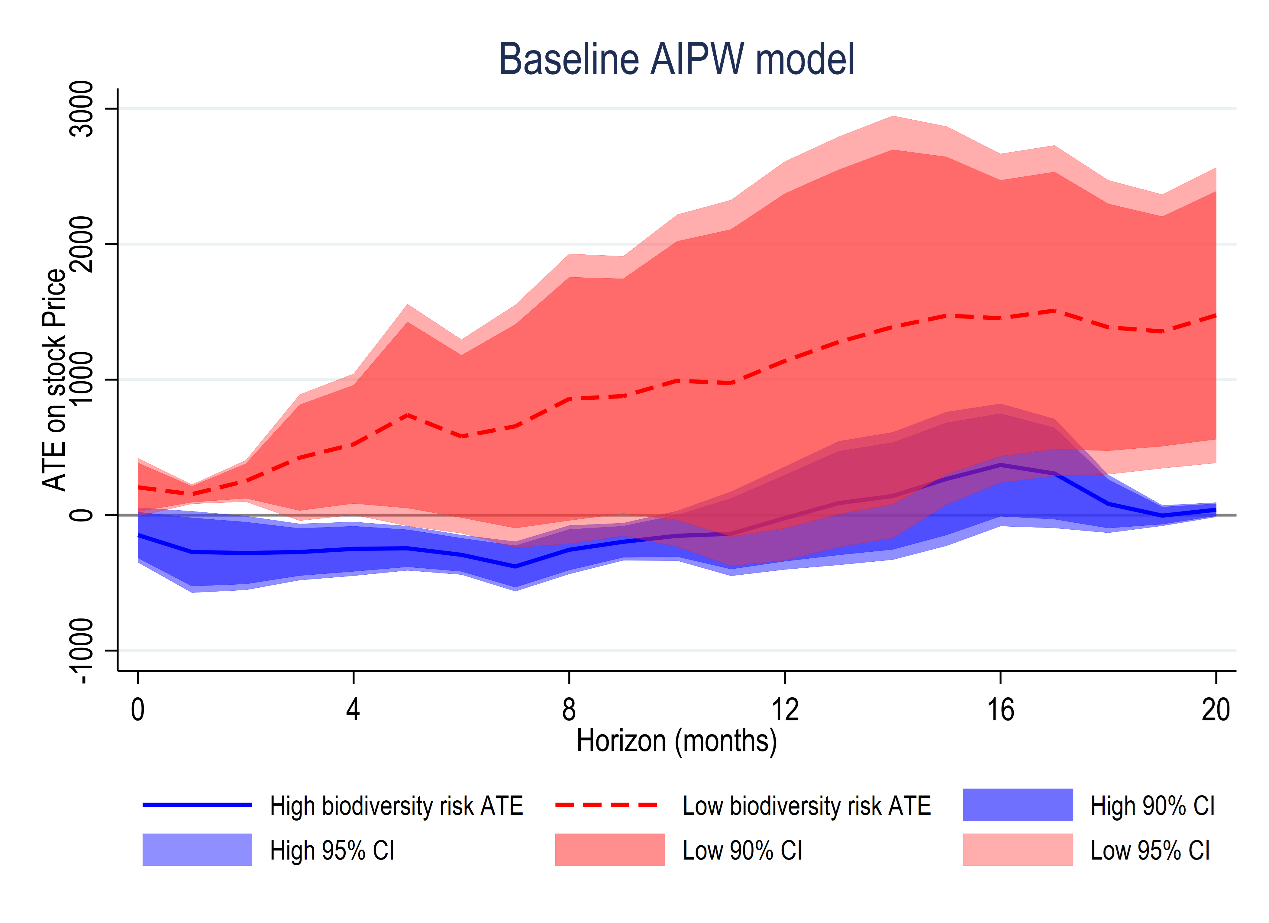}
\caption{Average Treatment Effects of Biodiversity Risk Shocks on Stock Prices.}
\label{fig:figure2}
\vspace{0.2cm}
\begin{minipage}{0.9\textwidth}
\footnotesize
\textit{Note:} Event-study estimated using the AIPW estimator of~\citet{robins1994},
implemented via Stata's \texttt{teffects aipw}. The treatment variable is the
standardized biodiversity risk indicator ($z\_bio$), classified into high-risk (top
quartile) and low-risk (bottom quartile) regimes relative to the middle group
(interquartile range). Both the propensity score and outcome models include the
one-month lag of the stock price index (\texttt{L1\_price}). Shaded areas represent
90\% and 95\% confidence intervals around the estimated ATEs at monthly horizons
$h = 0, \ldots, 20$.
\end{minipage}
\end{figure}

Figure~\ref{fig:figure2} reports the average treatment effects of biodiversity risk shocks on stock prices across horizons. The effects are strongly asymmetric. High biodiversity risk episodes are associated with declines of roughly 300--400 index points, concentrated in the medium term. In contrast, low biodiversity risk episodes are associated with higher stock price levels, with effects increasing over time and remaining positive throughout most horizons, though with wider confidence intervals. The results therefore suggest an asymmetric market response, with positive biodiversity-related news exerting a stronger impact on valuations than negative biodiversity-related news. This pattern is consistent with~\citet{ma2024biodiversity}, who attribute the negative effect to biodiversity degradation concerns repricing stocks associated with greater ecological exposure. The stronger positive response during low-risk periods may additionally reflect a relief effect, whereby the absence of biodiversity stress signals improved ecosystem stability and reduces the risk premium demanded by investors.

\subsection{Robustness}

We assess robustness along two dimensions. First, we run Granger causality in
quantiles. Appendix~\ref{app:A1} reports the p-values associated with the null
hypothesis of no causality from biodiversity risk to stock prices at different
quantiles $\tau$. The results reveal that biodiversity risk significantly affects
stock prices at the lower quantiles (0.05--0.25) and the upper quantiles
(0.60--0.95), while the relationship is not statistically significant around the
middle quantiles (0.30--0.55). This suggests that biodiversity risk mainly affects
stock prices during extreme market conditions, consistent with a nonlinear,
distribution-dependent relationship.

Second, we re-estimate the AIPW event study under three alternative specifications:
(i)~adding the one-month lag of the Euro Stoxx~50 index to control for broader
European equity market conditions; (ii)~adding the one-month lag of the Economic
Policy Uncertainty (EPU) index; and (iii)~including both additional controls
simultaneously. Appendix~\ref{app:A2} presents the results under the Euro Stoxx~50
specification. All four specifications produce qualitatively similar results,
suggesting that the estimated biodiversity effects are robust to the inclusion of
additional macro-financial control variables.

% =========================
% CONCLUSION
% =========================
\section{Conclusion}
\label{sectionConc}

This paper introduces novel text-based measures of biodiversity risk for four major
European countries and examines their impact on financial markets. Using panel Granger
causality and an AIPW event-study design, we demonstrate that biodiversity risk
exhibits a significant negative effect on stock prices, concentrated in the tails of
the return distribution and peaking at medium horizons of 3--12 months.

Our findings provide compelling evidence that biodiversity loss is not only an
ecological crisis but also a financial one. Investors appear to reprice risk in
response to biodiversity-related news, particularly during episodes of heightened
ecological concern. Our biodiversity indicators can therefore support early warning
systems for financial institutions and contribute to ongoing efforts to mainstream
nature-related risk disclosures. Incorporating biodiversity metrics into investment
strategies and regulatory stress-testing frameworks will become increasingly essential
as the EU Taxonomy and TNFD frameworks mature.

% =========================
% REFERENCES
% =========================
\clearpage
\bibliographystyle{chicago}
\bibliography{references}

% =========================
% APPENDIX
% =========================
\clearpage
\section*{Appendix}
\addcontentsline{toc}{section}{Appendix}
\renewcommand{\thesubsection}{\Alph{subsection}}
\setcounter{subsection}{0}
\renewcommand{\thefigure}{\Alph{subsection}.\arabic{figure}}
\setcounter{figure}{0}
\renewcommand{\thetable}{\Alph{subsection}.\arabic{table}}
\setcounter{table}{0}

% --- Appendix A ---
\subsection*{Appendix A: Data Collection Details}
\setcounter{subsection}{1}
\setcounter{figure}{0}
\addcontentsline{toc}{subsection}{Appendix A: Data Collection Details}
\label{app:data}

\textbf{Sampling strategy.} For computational efficiency and representative sampling,
we query 48 time snapshots per month: three days per month (5th, 15th, 25th) at four
times daily (00:00, 06:00, 12:00, 18:00 UTC), yielding approximately 1,000--5,000
articles per country-month.

\textbf{Keyword selection.} Following established word2vec methodologies for
identifying semantically-related terms~\citep{sautner2023, mikolov2013}, we employ
100+ biodiversity keywords derived from semantic similarity analysis in environmental
science literature. The keyword taxonomy comprises four categories: (1)~\textit{Core
biodiversity concepts}: biodiversity, biological diversity, biodiversity conservation,
marine biodiversity, ecosystem services, genetic diversity; (2)~\textit{Ecosystem
types and habitats}: ecosystems (marine, freshwater, forest, wetland, coastal),
tropical forests, coral reefs, rainforests, mangroves, seagrass meadows, protected
areas, national parks, biosphere reserves; (3)~\textit{Species and conservation}:
species, fauna, flora, wildlife, endangered species, extinction, endemic species,
migratory species, pollinators, apex predators; (4)~\textit{Threats and degradation}:
deforestation, habitat fragmentation, habitat loss, desertification, overexploitation,
overfishing, environmental degradation, soil erosion, invasive species.

\textbf{Geographic filtering.} For each country $c \in \{\text{France, Germany,
Italy, Spain}\}$, we implement two-stage filtering. \textit{Stage~1 --- Geographic
identification}: We search GDELT metadata fields (SourceCommonName, DocumentIdentifier
URL, V2Themes) for country-specific geographic terms (e.g., for France: \textit{France,
French, Paris, Marseille, Lyon, Corsica, French Guiana, R\'{e}union}; similar
comprehensive lists cover the remaining countries). \textit{Stage~2 --- Biodiversity
content filtering}: Among geographically-identified articles, we retain only those
containing at least one biodiversity keyword, ensuring that articles discuss
biodiversity issues specifically related to the focal country.

% --- Appendix B ---
\subsection*{Appendix B: Quantile Granger Causality Results}
\setcounter{subsection}{2}
\setcounter{figure}{0}
\addcontentsline{toc}{subsection}{Appendix B: Quantile Granger Causality Results}
\label{app:A1}

\begin{figure}[!htbp]
    \centering
    \includegraphics[width=\textwidth]{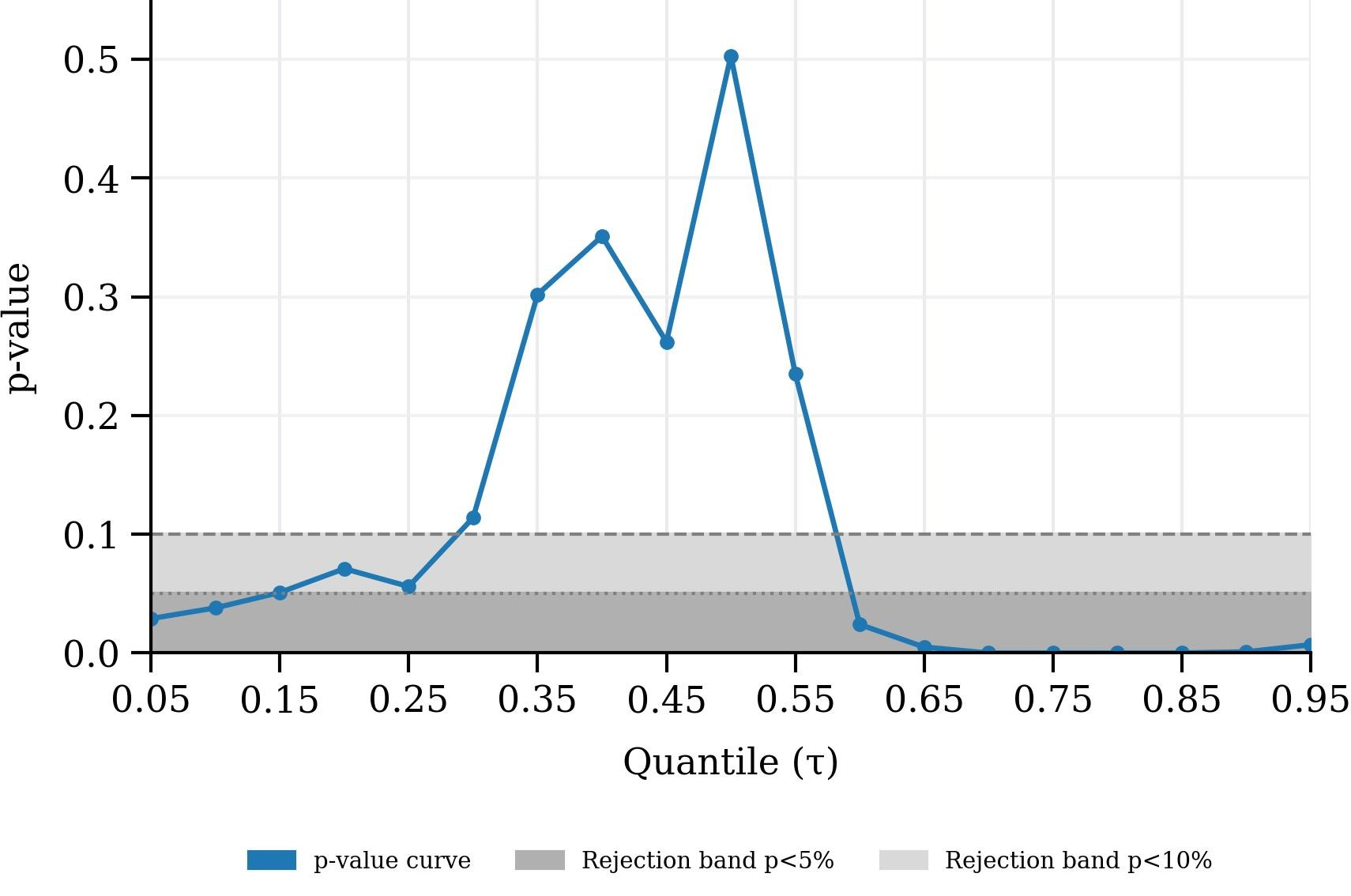}
    \caption{Quantile Granger Causality Test --- Panel EU4. \textit{Note:} P-values
    for the null hypothesis of no Granger causality from biodiversity risk to stock
    prices at each quantile $\tau$. Dashed lines indicate 5\% and 10\% significance
    thresholds.}
\end{figure}

% --- Appendix C ---
\subsection*{Appendix C: AIPW Event Study --- Robustness Results}
\setcounter{subsection}{3}
\setcounter{figure}{0}
\addcontentsline{toc}{subsection}{Appendix C: AIPW Event Study --- Robustness Results}
\label{app:A2}

\begin{figure}[!htbp]
    \centering
    \includegraphics[width=\textwidth]{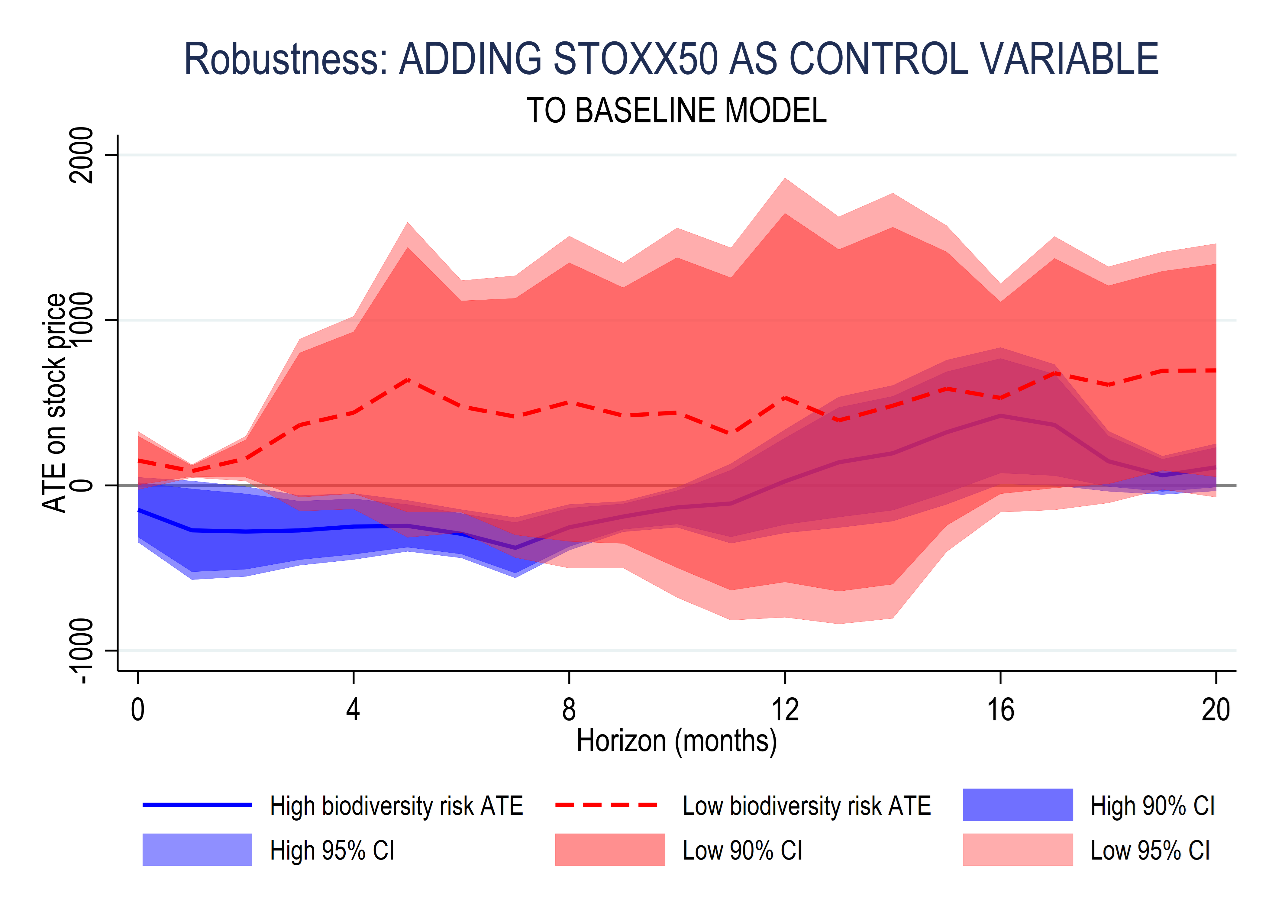}
    \caption{AIPW Event Study --- Robustness with Euro Stoxx~50 Control.
    \textit{Note:} Specification augments the baseline model with the one-month lag
    of the Euro Stoxx~50 index (\texttt{L1\_stoxx50}) in both the propensity score
    and outcome regression. High-risk (blue) and low-risk (red) regimes relative to
    the middle group. Shaded areas represent 90\% and 95\% confidence intervals.
    Results are qualitatively unchanged when additionally controlling for the
    one-month lag of the EPU index.}
\end{figure}

\begin{figure}[!htbp]
    \centering
    \includegraphics[width=\textwidth]{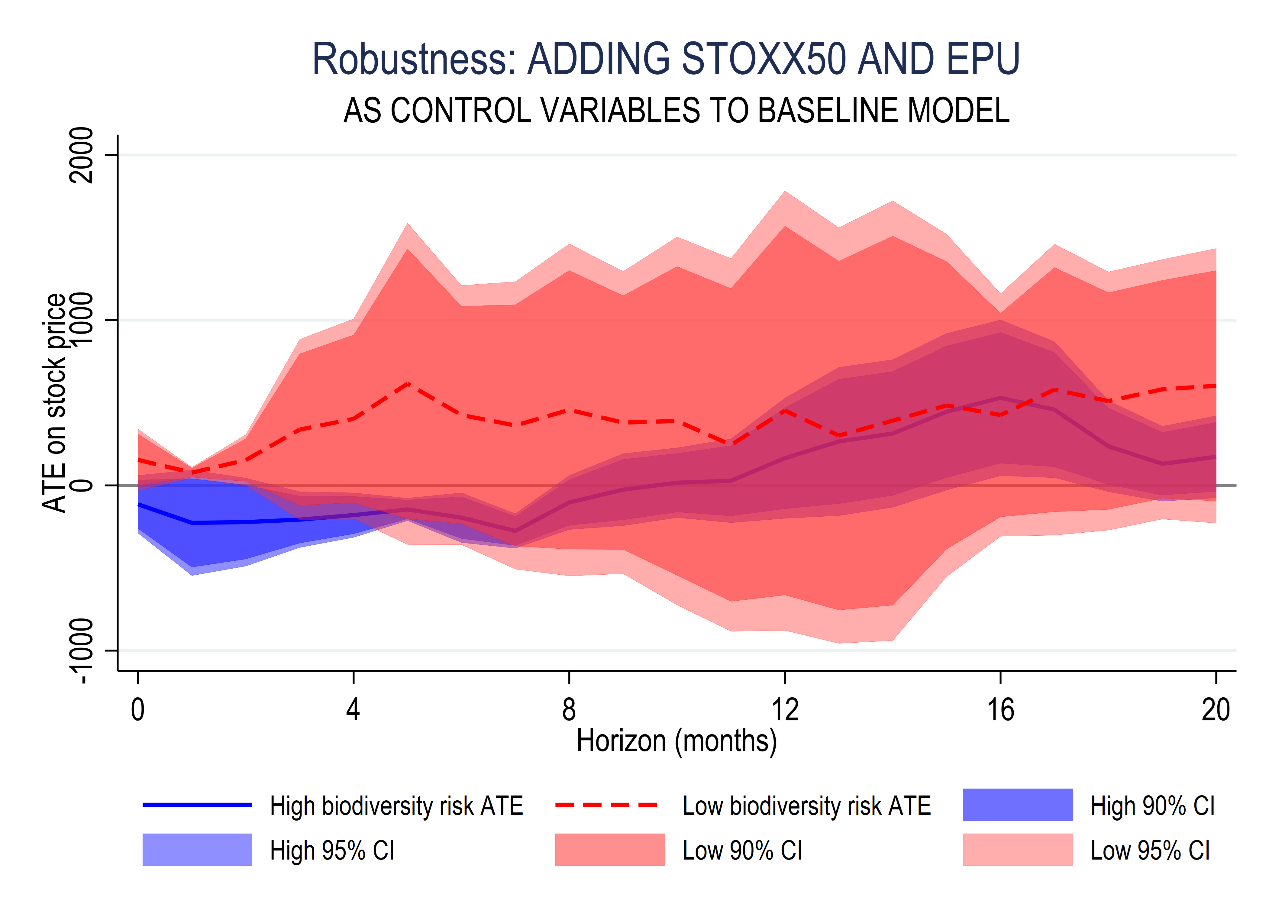}
    \caption{AIPW Event Study --- Robustness with Euro Stoxx~50 and EPU Control.
    \textit{Note:} Specification augments the baseline model with the one-month lag
    of the Euro Stoxx~50 index (\texttt{L1\_stoxx50}) in both the propensity score
    and outcome regression. High-risk (blue) and low-risk (red) regimes relative to
    the middle group. Shaded areas represent 90\% and 95\% confidence intervals.
    Results are qualitatively unchanged when additionally controlling for the
    one-month lag of the EPU index.}
\end{figure}

\begin{figure}[!htbp]
    \centering
    \includegraphics[width=\textwidth]{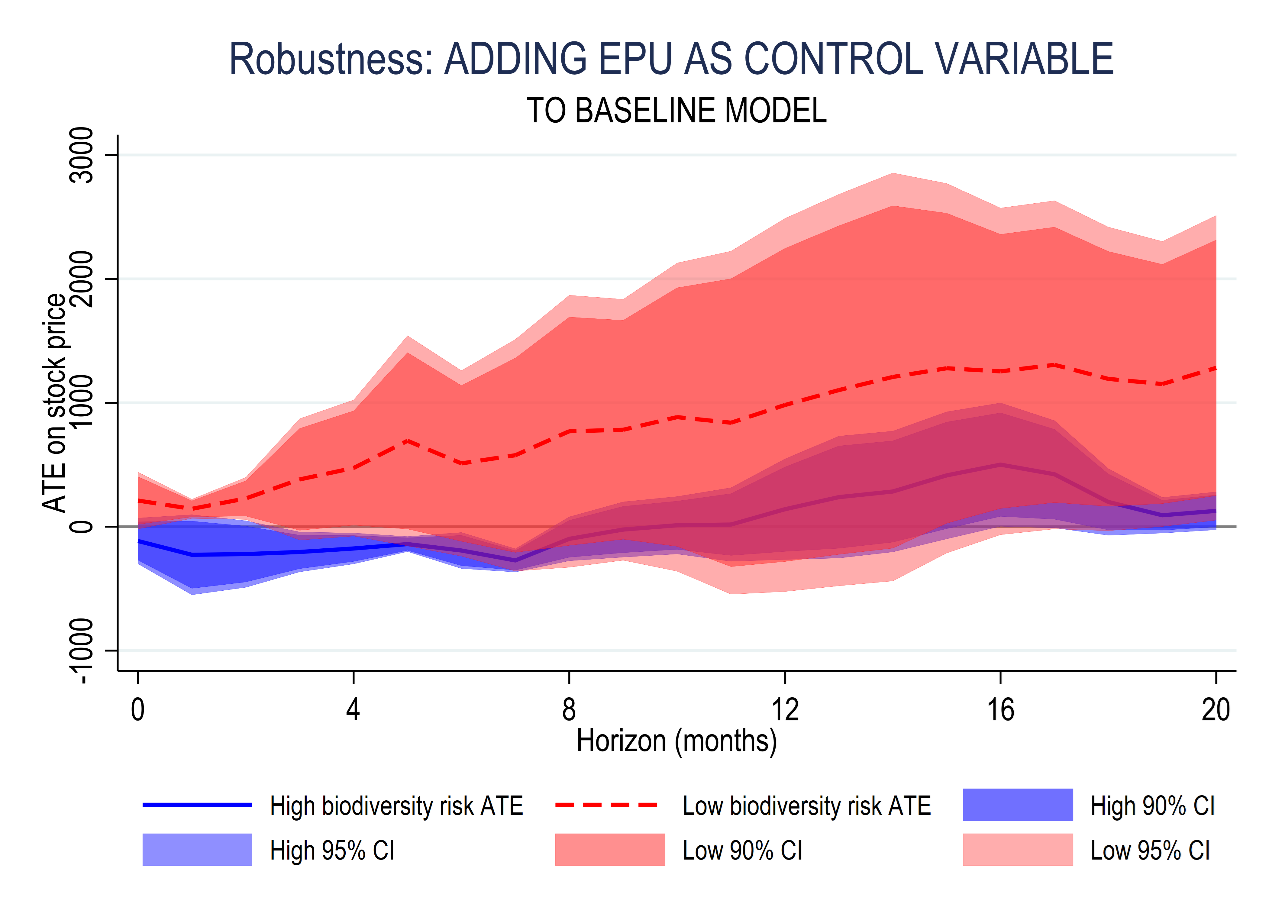}
    \caption{AIPW Event Study --- Robustness with EPU Control.
    \textit{Note:} Specification augments the baseline model with the one-month lag
    of the Euro Stoxx~50 index (\texttt{L1\_stoxx50}) in both the propensity score
    and outcome regression. High-risk (blue) and low-risk (red) regimes relative to
    the middle group. Shaded areas represent 90\% and 95\% confidence intervals.
    Results are qualitatively unchanged when additionally controlling for the
    one-month lag of the EPU index.}
\end{figure}

\end{document}